\documentclass[conference]{IEEEtran}
\IEEEoverridecommandlockouts
\usepackage{cite}
\usepackage{amsmath,amssymb,amsfonts}
\usepackage{algorithmic}
\usepackage{graphicx}
\usepackage{textcomp}
\usepackage{xcolor}
\usepackage{tikz}
\usepackage{pgfplots}
\pgfplotsset{compat=1.17}

\usepgfplotslibrary{statistics,groupplots}

\def\BibTeX{{\rm B\kern-.05em{\sc i\kern-.025em b}\kern-.08em
    T\kern-.1667em\lower.7ex\hbox{E}\kern-.125emX}}

\graphicspath{{../PICS/}}

\begin{document}

\title{RIS-Assisted 3D Spherical Splatting for Object Composition Visualization using Detection Transformers}

\bstctlcite{IEEEexample:BSTcontrol}

\author{
    \IEEEauthorblockN{
        Anastasios T. Sotiropoulos\IEEEauthorrefmark{1}, 
        Stavros Tsimpoukis\IEEEauthorrefmark{2}, 
        Dimitrios Tyrovolas\IEEEauthorrefmark{1}\IEEEauthorrefmark{3},  \\ 
        Sotiris Ioannidis\IEEEauthorrefmark{3}\IEEEauthorrefmark{4},
        Panagiotis D. Diamantoulakis\IEEEauthorrefmark{1},
        George K. Karagiannidis\IEEEauthorrefmark{1}, and
        Christos K. Liaskos\IEEEauthorrefmark{2}\IEEEauthorrefmark{5}
    }
    
    \IEEEauthorblockA{
        \IEEEauthorrefmark{1}Department of Electrical and Computer Engineering, Aristotle University of Thessaloniki, Thessaloniki, Greece,
    }
        
    \IEEEauthorblockA{
        e-mail: \{anastsotir, tyrovolas, padiaman, geokarag\}@ece.auth.gr
    }
    
    \IEEEauthorblockA{
        \IEEEauthorrefmark{2}Department of Computer Science and Engineering, University of Ioannina, Ioannina, Greece,
    }
    
    \IEEEauthorblockA{
        e-mail: \{s.tsimpoukis, cliaskos\}@uoi.gr
    }

    \IEEEauthorblockA{
        \IEEEauthorrefmark{3}Dienekes SI IKE, 71414 Heraklion, Greece
    }
    
    \IEEEauthorblockA{
        \IEEEauthorrefmark{4} Department of Electrical and Computer Engineering, Technical University of Crete, Chania, Greece,
    }
    
    \IEEEauthorblockA{
        e-mail: sotiris@ece.tuc.gr
    }
    
    \IEEEauthorblockA{
        \IEEEauthorrefmark{5}Institute of Computer Science, Foundation for Research and Technology Hellas, Greece.
    }
    
}

\maketitle

\begin{abstract}
The pursuit of immersive and structurally aware multimedia experiences has intensified interest in sensing modalities that reconstruct objects beyond the limits of visible light. Conventional optical pipelines degrade under occlusion or low illumination, motivating the use of radio-frequency (RF) sensing, whose electromagnetic waves penetrate materials and encode both geometric and compositional information. Yet, uncontrolled multipath propagation restricts reconstruction accuracy. Recent advances in Programmable Wireless Environments (PWEs) mitigate this limitation by enabling software-defined manipulation of propagation through Reconfigurable Intelligent Surfaces (RISs), thereby providing controllable illumination diversity. Building on this capability, this work introduces a PWE-driven RF framework for three-dimensional object reconstruction using material-aware spherical primitives. The proposed approach combines RIS-enabled field synthesis with a Detection Transformer (DETR) that infers spatial and material parameters directly from extracted RF features. Simulation results confirm the framework’s ability to approximate object geometries and classify material composition with an overall accuracy of 79.35\%, marking an initial step toward programmable and physically grounded RF-based 3D object composition visualization.
\end{abstract}


\section{Introduction} \label{S: Intro3}
Driven by the growing demand for immersive and interactive experiences, modern multimedia systems increasingly depend on sensing technologies capable of reconstructing objects and environments with structural precision~\cite{siriwardhana2021survey}. Among these, optical pipelines have long dominated scene perception, yet their reliance on direct visibility and controlled illumination makes them unreliable under occlusion or low-light conditions, while multi-camera setups introduce substantial synchronization complexity. These limitations have shifted attention toward radio-frequency (RF) sensing, which leverages the ability of electromagnetic (EM) waves to penetrate materials and capture both geometric and compositional information beyond the reach of visible light~\cite{khunteta2022rf}. Despite these advantages, RF perception in indoor environments remains hindered by the random and diffuse nature of multipath propagation, which obscures spatial relationships within the received field. Recent advances in Programmable Wireless Environments (PWEs) address this limitation by introducing deterministic control over the propagation process through Reconfigurable Intelligent Surfaces (RISs)~\cite{liaskos2020internet}. By dynamically steering, focusing, or diffusing incident waves under software-defined control, PWEs transform the environment from a passive transmission medium into an active, reconfigurable component of the sensing system, establishing a solid foundation for controlled RF-based multimedia services.

The growing capability to program EM propagation through RISs has opened new directions for representing the physical interaction between waves and their surrounding environment in a structured and interpretable manner. Among recent advances in scene representation, the 3D Gaussian Splatting (3DGS) paradigm~\cite{3D_Gaussian_Splatting} has shown remarkable effectiveness in the optical domain by modeling complex environments as assemblies of spatially distributed ellipsoidal primitives, each defined by geometric and material appearance parameters. Through differentiable optimization, these primitives collectively reconstruct the scene’s radiance field, achieving photorealistic rendering and compact modeling of both structure and appearance. In more detail, the decomposition of intricate scenes into localized volumetric elements in 3DGS naturally aligns with the goals of RF perception, where EM interactions inherently encode information about geometry and material composition. 
Existing efforts to connect RF propagation with primitive-based modeling have remained limited to radiance-field reconstruction and relied on prior optical information from camera-derived point clouds rather than directly recovering surfaces from RF alone~\cite{wrfgs}. In \cite{3D_Object_Reconstruction_ROBOTS_RF_ADIB_FADEL_MIT}, object surfaces are reconstructed by estimating a dense per-voxel surface-normal field from multi-view mmWave measurements, however, this approach does not infer or model the objects’ material properties. To the best of the authors’ knowledge, a comprehensive formulation of a material-aware, primitive-based object representation operating directly within the RF domain has yet to be established.

In this direction, the present work introduces a programmable RF sensing framework that operates within a PWE to visualize and reconstruct three-dimensional objects through a novel 3D Spherical Splatting formulation. The approach represents each object as a material-aware assembly of spherical primitives, leveraging the controlled illumination diversity enabled by RISs, where beam steering and diffuse scattering are jointly programmed to enrich the measured wavefronts with discriminative geometric and EM information. These observations are subsequently interpreted by a Detection Transformer (DETR) trained to infer the spatial configuration and material attributes of the constituent spheres directly from the extracted RF features. To enhance physical interpretability while maintaining expressive modeling capacity, spherical primitives are introduced as an alternative to anisotropic Gaussians, providing a simpler yet effective volumetric representation tailored to RF propagation. Consequently, the proposed framework establishes an RF-domain extension of 3D Gaussian Splatting, in which programmable propagation and deep learning cooperate to achieve controllable, material-aware object composition visualization within PWEs.

\section{System Model} \label{S:system_model}

\begin{figure}
    \centering
    \includegraphics[width=0.8\linewidth]{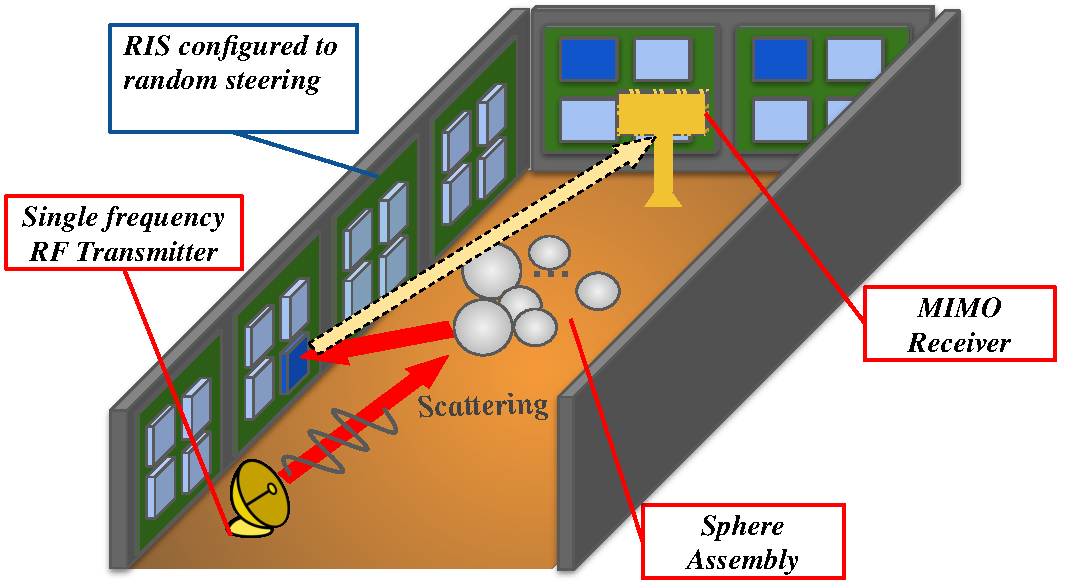}
    \vspace{-8pt}
    \caption{Overview of the 3D-object reconstruction system.}
    \label{fig:system_model}
\end{figure}

We consider an indoor PWE, illustrated in Fig.~\ref{fig:system_model}, comprising a rectangular room with RIS-coated walls, a single-antenna transmitter (Tx), and a multiple-input multiple-output (MIMO) receiver (Rx). The Tx emits a continuous single-tone RF signal at carrier frequency $f_c$, which propagates through the environment and interacts with the RIS surfaces and any object placed within the sensing region. Each RIS element applies a controllable phase shift to the incident wave, enabling programmable manipulation of the multipath propagation. The reflection behavior of the RIS array is determined by an offline-constructed codebook~\cite{RIS_Codebook}, denoted as $\mathcal{C} = \{\boldsymbol{\Theta}_1, \boldsymbol{\Theta}_2, \ldots, \boldsymbol{\Theta}_c\}$, where each $\boldsymbol{\Theta}_c$ defines a distinct steering configuration. During operation, the PWE sequentially adopts different states from $\mathcal{C}$, producing unique propagation signatures under single-frequency illumination. These controlled variations allow the environment to act as an active operator of the RF channel, generating distinct scattering responses that encode the object’s geometric and EM characteristics. Consequently, the Rx array observes a series of PWE-shaped wavefronts, from which the object’s three-dimensional structure and material composition are inferred directly from its measured RF scattering profile.

The measured scattering profile within the PWE is dictated by the object's EM distribution, described by a continuous spatial complex permittivity function $\varepsilon(\boldsymbol{r})$. Since directly resolving this volumetric field is intractable in practice, the illuminated region is approximated through finite elementary volumes with uniform complex permittivity values, providing a compact parametric representation of the object. In the considered system model, these volumes are modeled as spherical primitives, each defined by its center coordinates $\boldsymbol{\psi}_s=(x_s,y_s,z_s)$, radius $r_s$, and complex permittivity $\varepsilon_s=\varepsilon'_s-j\varepsilon''_s$, where $s\in\{1,2,\ldots,S\}$. Under a given RIS configuration $\boldsymbol{\Theta}_c$, the transmitted signal undergoes multiple reflections and scattering events involving both the RIS-coated walls and the set of spherical primitives. Consequently, the baseband signal received at the $n$-th antenna of the Rx can be expressed as
\begin{equation}
\small
r_n^{(c)} = \sum_{s=1}^{S}\sum_{k=1}^{K_s} \alpha_{s,k}^{(c)} 
e^{-j\frac{2\pi f_c}{c_0}d_{s,k}^{(c)}} + w_n,
\end{equation}
where $\alpha_{s,k}^{(c)}$ and $d_{s,k}^{(c)}$ denote the complex attenuation coefficient and propagation distance of the $k$-th multipath component associated with the $s$-th primitive, $c_0$ is the speed of light, and $w_n$ represents additive white Gaussian noise with zero mean and variance $\sigma^2$. The received wavefronts across all $N_\mathrm{r}$ antennas, represented as $\mathbf{r}^{(c)}=[r_1^{(c)},r_2^{(c)},\ldots,r_{N_\mathrm{r}}^{(c)}]^{\top}$,  capture the composite scattering behavior of the environment under configuration $\boldsymbol{\Theta}_c$, forming the observation space from which the object’s geometry and material are inferred.

\section{Methodology} \label{S: Alt Methodology}

Building upon the described system model, the methodology establishes the process through which the PWE and the machine learning (ML) framework jointly enable object inference from measured RF wavefronts. The overall approach follows an RF implementation of the 3DGS concept used in optical scene synthesis, where a scene is decomposed into simple volumetric elements that collectively describe its structure. In the RF domain, this analogy translates into reconstructing an object’s geometry and material composition through programmable field observations that reveal its scattering behavior under different illumination conditions. Within this framework, the PWE provides controllable multipath diversity, exposing complementary perspectives of the same object, while the ML component learns to interpret these responses to recover the object’s parametric representation. As our goal is to enable the system to reconstruct objects using spherical primitives of arbitrary size and material, we require appropriately curated training data. To this end, we generate a synthetic dataset by placing spheres at random locations via Bridson’s Poisson-disk sampling \cite{Bridson_Algorithm}, consistent with the studied PWE scenario. We enforce a minimum separation to avoid significant overlaps, preventing degenerate scenes. The dataset comprises multiple scene realizations that pair each sphere’s center coordinates, radii, and material labels with simulated RF wavefronts, from which the corresponding feature vectors are derived. The following subsections detail the steps of this process, beginning with the extraction of physically meaningful RF features from the measured wavefronts and continuing with the deep-learning model that performs the final inference.



\subsection{Feature Extraction}\label{S:meth_feature}

During training, the PWE sequentially activates a finite set of configurations from a predefined RIS codebook, each producing a distinct EM illumination condition within the sensing region.
Under every configuration, a unique RF wavefront is captured, embedding the object’s geometric and EM characteristics into the measured field. These wavefronts are subsequently transformed into a compact RF signature that serves as input to the learning module.
The resulting feature representation combines EM and communication-inspired quantities that describe the propagation behavior observed across the receiver array. Formally, the input feature vector is defined as $\mathbf{x} = \bigoplus_{c=1}^{C}\Phi_c(\mathcal{W})$,
where $\mathbf{x}$ denotes the concatenated feature representation, $\mathcal{W}$ represents the measured RF wavefront data, $c \in \mathcal{C}$ indexes the active PWE configuration, and $\Phi_c(\cdot): \mathcal{W} \!\to\! \mathbb{C}^{N_{\mathrm{r}} \times N_f}$ is the feature-extraction operator, with $N_{\mathrm{r}}$ being the number of receiver antennas and $N_f$ the number of extracted features per antenna. In this direction, the following subsections present the extracted quantities, which include polarization-aware power and phase attributes, angular dispersion measures, and temporal delay statistics that collectively describe the multipath propagation profile.

\subsubsection{Polarization-Aware Power and Phase Features} The polarization state of the received field conveys information about the object’s EM composition and surface orientation, as variations in material or geometry modulate the relative levels of the co-polarized and cross-polarized backscatter ratio of the reflected components.
Capturing this behavior enriches the description of the scattering process beyond power-only analysis and facilitates discrimination among materials with similar geometric responses.
In the considered scenario, the field observed at each antenna results from the superposition of multiple propagation paths, each contributing to the horizontally and vertically polarized components of the received wave. In this direction, the complex response of the $k$-th propagation path between transmit and receive polarization states $q,p\in\{h,v\}$ can be expressed as $h_{pq}(n,k)=\left|h_{pq}(n,k)\right|\,e^{j\phi_{pq}(n,k)}$, where $\phi_{pq}(n,k)=\arg\!\big(h_{pq}(n,k)\big)$ and $\arg(\cdot)$ denoting the argument of a complex number.By summing all paths, the total received field per polarization becomes $X_{p,n}=X_{\mathrm{co}}(n,p)+X_{\mathrm{xpol}}(n,p)$, where $X_{\mathrm{co}}(n,p)$ and $X_{\mathrm{xpol}}(n,p)$ aggregate the co-polarized and cross-polarized contributions corresponding to polarization $p$. Accordingly, the polarization-dependent features are defined as $P_{p,n}=\lVert X_{p,n}\rVert^2$ and $\omega_{p,n}=\arg\!\big(X_{p,n}\big)$, which jointly characterize the object-induced polarization response within the PWE and form a fundamental component of the extracted feature set.

\subsubsection{Power-Weighted Equivalent Angle-of-Arrival Features}\label{S:meth_aoa} While polarization features describe variations in the EM response of the object, additional spatial information arises from the directions of the multipath components reaching the receiver array. The distribution of these arrival directions reflects both the environment’s geometry and the object’s structural shape, motivating the inclusion of an angle-of-arrival (AoA) descriptor within the feature set. Therefore, for each receiver antenna, the received signal comprises multiple propagation copies arriving from azimuth–elevation angles $(\phi_{n,k},\theta_{n,k})$, each carrying horizontal and vertical field components $E_{h}(n,k)$ and $E_{v}(n,k)$. The power of the $k$-th copy is expressed as $I_{n,k} = \lVert E_{h}(n,k) \rVert^{2} + \lVert E_{v}(n,k) \rVert^{2}$. To summarize the spatial distribution of these multipath arrivals into a representative direction per antenna, a power-weighted aggregation is introduced. In more detail, each propagation path is modeled as a three-dimensional vector whose magnitude reflects its received power and whose orientation corresponds to its arrival direction, expressed as $\boldsymbol{r}_{n,k} = I_{n,k}\,\hat{\boldsymbol{r}}_{n,k}$, with $\hat{\boldsymbol{r}}_{n,k}=[\cos\theta_{n,k}\cos\phi_{n,k},\,\cos\theta_{n,k}\sin\phi_{n,k},\,\sin\theta_{n,k}]^{\top}$ denoting the unit vector associated with the azimuth–elevation pair $(\phi_{n,k},\theta_{n,k})$. Therefore, summing all power-weighted vectors produces a resultant whose orientation defines the equivalent AoA perceived by the $n$-th antenna. The aggregate azimuth and elevation angles follow from the Cartesian components of $\boldsymbol{r}_{n,k}$ as
\begin{equation}
\small
    \phi_{t,n} = \arctan2\!\left(\sum_k y_{n,k},\, \sum_k x_{n,k}\right),
\end{equation}
and
\begin{equation}
\small
    \theta_{t,n} = \arcsin\!\left(\frac{\sum_k z_{n,k}}{\big\lVert \big(\sum_k x_{n,k},\,\sum_k y_{n,k},\,\sum_k z_{n,k}\big) \big\rVert}\right),
\end{equation}
where $x_{n,k}$, $y_{n,k}$, and $z_{n,k}$ denote the Cartesian components of $\boldsymbol{r}_{n,k}$, and $\arctan2(\cdot,\cdot)$ is the four-quadrant inverse tangent function. Thus, the pair $(\phi_{t,n},\theta_{t,n})$ encapsulates the dominant direction of multipath arrivals observed at antenna $n$, capturing the spatial characteristics of the object’s scattering within the PWE.

\subsubsection{Angular RMS Spread Features} Beyond the dominant angle of arrival, the spatial dispersion of multipath components also carries valuable information about the environment and the object’s geometry. A tightly clustered distribution of arrival directions indicates a dominant reflection or strong specular path, whereas a broader spread suggests increased scattering or structural complexity. To quantify this behavior, the angular root-mean-square (RMS) spread is introduced as a measure of the directional variance observed at each antenna. In more detail, for receiver antenna $n$, the power-weighted RMS angular spread is defined as
\begin{equation}
\small
    \sigma_{\psi,n} = 
    \sqrt{
    \frac{
        \sum_k I_{n,k}\left(\psi_{n,k} - \psi_{t,n}\right)^2
    }{
        \sum_k I_{n,k}
    }},
\end{equation}
where $\psi \in \{\phi, \theta \}$. Thus, the pair $(\sigma_{\phi,n},\sigma_{\theta,n})$ captures the extent of multipath dispersion in two angular dimensions, complementing the equivalent AoA feature by describing the spread rather than the mean direction of the received energy.

\subsubsection{Delay Statistics Features} In addition to spatial dispersion, the temporal distribution of multipath components provides a complementary source of information regarding the propagation behavior within the PWE.
As the wavefronts follow paths of different lengths and reflection orders, their arrivals at the receiver occur over a range of delays rather than simultaneously, forming a characteristic temporal profile. This delay structure reflects the complexity of the propagation environment, as increased scattering or material diversity introduces broader temporal spreading. To quantify this behavior, two complementary descriptors are employed that jointly capture the central tendency and the temporal extent of the received energy. The first descriptor, the power-weighted mean excess delay, expresses the average arrival time of the signal energy and is given by
\begin{equation}
\small
    \tau_{\mathrm{av},n} =
    \frac{\sum_k I_{n,k}\tau_{n,k}}{\sum_k I_{n,k}}.
\end{equation}
Additionally, the RMS delay spread, measures the temporal variance of  arrivals around their mean and is expressed as
\begin{equation}
\small
    \sigma_{\tau,n} =
    \sqrt{
        \frac{\sum_k I_{n,k}\tau_{n,k}^2}{\sum_k I_{n,k}}
        - \tau_{\mathrm{av},n}^2
    }.
\end{equation}
Together, $\tau_{\mathrm{av},n}$ and $\sigma_{\tau,n}$ form the temporal counterpart of the angular features, providing a complementary view of how the environment and the object collectively shape the received waveform within the PWE.

\subsection{Detection Transormer Modeling}
Having established a compact feature representation that encodes the spatial, temporal, and polarization characteristics of the RF wavefronts, the next step concerns recovering the object’s physical parameters. The extracted features carry rich EM information, yet their relationship to geometry and material composition stems from nonlinear scattering effects that precludes closed-form analytical modeling. Consequently, the inference process is cast as a data-driven regression task for spheres geometrical attributes and classification task for material classification and the goal becomes estimating a configuration of spherical primitives that best reproduces the observed propagation behavior. Using our synthetic dataset, the network is trained to map measurable RF signatures to the underlying physical parameters, thereby establishing a reproducible, data-driven bridge between the EM and geometric domains.

Building upon this formulation, the learning problem must now reflect the structural nature of the data. Each training instance corresponds to a collection of spherical primitives whose ordering bears no inherent meaning. This set-based property naturally aligns with the paradigm of three-dimensional object detection, where each primitive constitutes an individual instance defined by its spatial position, radius, and material descriptor. To process such data effectively, the network must reason jointly across primitives while preserving spatial coherence within the learned features. These capabilities are intrinsically supported by detection transformer architectures. Within this framework, DETR~\cite{carion2020end} provides a particularly suitable foundation, as its design couples global attention with permutation-invariant decoding, enabling set prediction over primitives while preserving geometric coherence across outputs. In our implementation, we omit the convolutional backbone and operate directly in our RF-derived feature space. Following the projection layer of the architecture, the transformer encoder ingests handcrafted feature vectors arranged as a feature map whose spatial axes align with antenna locations and whose channel dimension encodes the feature components, enabling the spatial reasoning needed for material aware volumetric reconstruction.

Each training sample pairs the extracted features with the ground truth parameters of its constituent spheres, casting the task as supervised set prediction over unordered primitives. Predictions are aligned to ground truth via optimal bipartite assignment computed by the Hungarian algorithm. Localization precision is promoted through a linear combination of $\ell_1$ loss, which penalizes absolute deviations between predicted and reference parameters and a 3D Generalized Intersection over Union (GIoU) loss adapted from \cite{Rezatofighi_2019_CVPR}, which measures the overlap between predicted and reference spheres while accounting for size and spatial misalignment. The 3D GIoU loss is modified in account of the 3D spherical geometry. For a predicted sphere $A$ and a reference sphere $B$ in $\mathbb{R}^3$, and their minimum enclosing sphere $C$, the adapted GIoU loss is given by
\begin{equation}
\small
\mathcal{L}_{GIoU} = IoU - \frac{|C \setminus (A \cup B)|}{|C|},
\end{equation}
where \( IoU = \frac{|A \cap B|}{|A \cup B|} \) and $|\cdot|$ denotes the volumetric operator. The individual sphere volumes are defined as $|A| = \tfrac{4}{3}\pi r_A^3$ and $|B| = \tfrac{4}{3}\pi r_B^3$, the intersection volume $|A \cap B|$ follows from the closed-form geometry of intersecting spheres, and $|C|$ is derived from the radius of the minimal enclosing sphere. This formulation preserves geometric interpretability throughout training, ensuring that network convergence remains consistent with the physical representation of the reconstructed object. Finally, classification performance is optimized using the negative log-likelihood function.

After training, the model outputs \(N\) queries, each with a corresponding predicted material-aware spherical primitive $\hat{\mathbf{p}}_i = (\hat{x}_i, \hat{y}_i, \hat{z}_i, \hat{r}_i)$ accompanied by probabilities for each material class including the empty material class (\(\varnothing\)). Collectively, these outputs define a probabilistic reconstruction of the scene, where each prediction represents a potential material component of the object with an associated degree of certainty. The predicted material label and confidence is selected as
\begin{equation}
\small
\hat{l}_i = \arg\max_{l \in \{1,\dots,L\}} p_i(l), 
\qquad
s_i = \max_{l \in \{1,\dots,L\}} p_i(l),
\end{equation}
where $\hat{l}_i$ is the predicted material label, $s_i$ the associated confidence, and $p_i(l)$ the probability for class label $l$.
To derive a consistent geometric representation, the predictions are filtered using a confidence threshold $\tau$, retaining only those primitives deemed reliable by the model. Formally, the resulting detection set is expressed as
\begin{equation}
\small
\hat{\mathcal{S}}_{\tau} = \left\{\, (\hat{\mathbf{p}}_i, \hat{y}_i, s_i) \ \big|\ s_i \ge \tau,\ i=1,\dots,N \,\right\}.
\end{equation}
The threshold $\tau$ effectively regulates the level of structural detail in the reconstruction, allowing the number of retained primitives to adapt to the object’s geometric complexity. Through this filtering mechanism, the learned model transforms the extracted RF features into an interpretable volumetric representation, completing the inference stage of the proposed framework.

\color{black}
  
\section{Simulation Results} \label{S: Alt Sims}

\begin{table}[t!]
  \caption{Simulation Parameters}
  \label{tab:sim_params}
  \centering
  \resizebox{0.95\linewidth}{!}{
  \begin{tabular}{llll}
    \hline\hline
    \textbf{Parameter} & \textbf{Value} & \textbf{Parameter} & \textbf{Value} \\
    \hline
    PWE Dimension & \(6.5\,\text{m} \times 10\,\text{m} \times 4\,\text{m}\) &  Frequency & \(2.8\,\text{GHz}\)\\
    
    RIS Dimensions & \(1\,\text{m} \times 1\,\text{m}\) &  Number of Spheres & \(12\)\\
    Tx-position & \(\left(2.75,\,4.5,\,3.5\right)\,\text{m}\) &  \(x_s\) & \(\left[-1.5,\,1.5\right]\,\text{m}\) \\

    Tx Power & \(0\,\text{dBm}\) &  \(y_s\) & \(\left[-3.5,\,3.5\right]\,\text{m}\)\\
    
    Rx-position & \(\left(0,\,-5,\,2\right)\,\text{m}\) &  \(z_s\) & \(\left[1.75,\,2.25\right]\,\text{m}\)\\
    Rx Dimensions & \(8 \times 8\) &  \(r_s\) & \(\left[0.25,\,0.5\right]\,\text{m}\)\\

    \hline\hline
  \end{tabular}
  }
\end{table}

\begin{table}[t!]
  \caption{DETR Implementation Parameters}
  \label{tab:DETR}
  \centering
  \resizebox{\linewidth}{!}{
  \begin{tabular}{llll}
    \hline\hline
    \textbf{Parameter} & \textbf{Value} & \textbf{Parameter} & \textbf{Value} \\
    \hline
    Dataset Size & $300,000$ & Attention heads & $8$ \\
    Train/test split & $80/20$ & Hidden dimensions & $64$ \\
    Epochs & $300$ & Feed-forward dimensions & $512$ \\
    Batch Size & $128$ & $N$ object queries & $64$ \\
    Encoder/ Decoder Layers & $4/4$ & $\tau$ & $52\%$ \\
    \hline\hline
  \end{tabular}
  }
\end{table}

\begin{table}[t!]
  \caption{Material EM properties}
  \label{tab:material_properties}
  \centering
  \begin{tabular}{lccccc}
    \hline
    Materials & brick & wood & glass & ceiling-board & metal \\
    \hline
    $\varepsilon'_s$ & 3.91 & 1.99 & 6.31 & 1.48 & 1.00 \\
    $\sigma$ & 0.028 & 0.014 & 0.014 & 0.003 & 10000000 \\
    \hline
  \end{tabular}
\end{table}

In this section, the performance of the proposed framework is evaluated through ray-tracing simulations designed to emulate the programmable multipath behavior of the PWE. The simulated setup employs dually polarized, isotropic illumination for both transmitter and receiver layouts, while each RIS surface operates under $5$ independent steering realizations. This randomized illumination introduces controlled variability across configurations, enriching the diversity of recorded RF wavefronts and improving the robustness of feature learning. EM propagation is modeled using the MATLAB ray tracer, which computes the corresponding wavefronts for each PWE realization in accordance with the feature-extraction principles of Section~\ref{S:meth_feature}. In this model, up to four reflections per ray are considered, and diffraction effects are neglected to isolate the impact of specular multipath components. The material composition of each spherical primitive is drawn from MATLAB internal library, generating training scenes populated with spheres of various materials, whose parameters are listed in Table~\ref{tab:material_properties}. The remaining simulation parameters and DETR implementation details are summarized in Tables~\ref{tab:sim_params} and~\ref{tab:DETR}, respectively.

\def\HSEP{20pt}   
\def\VSEP{7.5pt}   
\def\AXW{2.5cm}   
\def\AXH{2.2cm}   

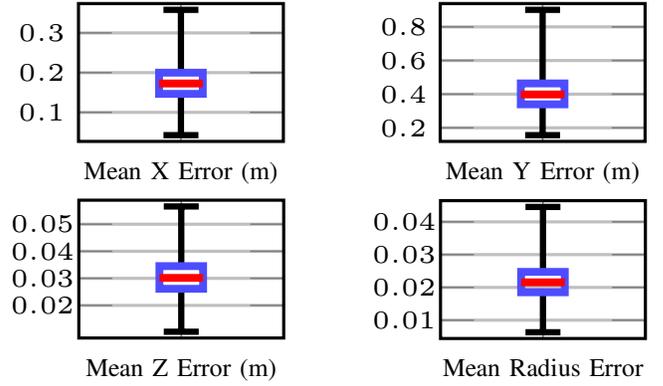
\begin{figure}[t!]
\centering
\resizebox{\columnwidth}{!}{%
\begin{tikzpicture}
\begin{groupplot}[
  group style={group size=2 by 2, horizontal sep=\HSEP, vertical sep=\VSEP},
  width=\AXW, height=\AXH,
  boxplot/draw direction=y,
  boxplot/box extend=0.32,
  xmin=0.5, xmax=1.5, xtick={1},
  enlarge x limits=0.25,
  ymajorgrids,
  scaled y ticks=false,                               
  tick label style={font=\tiny},
  xticklabel style={font=\tiny, scale=0.65, yshift=-0.1ex}, 
  yticklabel style={/pgf/number format/fixed,         
                    /pgf/number format/precision=2,
                    font=\tiny, scale=0.65, xshift=0.2ex},
  boxplot/every box/.style={draw=blue!70, line width=1pt},
  boxplot/every whisker/.style={black, line width=0.8pt},
  boxplot/every median/.style={red, line width=1pt},
  boxplot/every outlier/.style={black, mark=*, mark size=1pt},
]

\nextgroupplot[
  xticklabels={Mean X Error (m)},
  ymin=0.0269, ymax=0.3739,
  ytick={0.10,0.20,0.30}           
]
\addplot+[boxplot prepared={
  median=0.172744, lower quartile=0.146604, upper quartile=0.199918,
  lower whisker=0.042659, upper whisker=0.358129
}] coordinates {};

\nextgroupplot[
  xticklabels={Mean Y Error (m)},
  ymin=0.1190, ymax=0.9386,
  ytick={0.20,0.40,0.60,0.80}
]
\addplot+[boxplot prepared={
  median=0.398120, lower quartile=0.339511, upper quartile=0.463851,
  lower whisker=0.156241, upper whisker=0.901310
}] coordinates {};

\nextgroupplot[
  xticklabels={Mean Z Error (m)},
  ymin=0.00795, ymax=0.05888,
  ytick={0.02,0.03,0.04,0.05},
  yticklabel style={/pgf/number format/fixed,
                    /pgf/number format/precision=3,  
                    font=\tiny, xshift=0.4ex}
]
\addplot+[boxplot prepared={
  median=0.030163, lower quartile=0.026139, upper quartile=0.034519,
  lower whisker=0.010269, upper whisker=0.056566
}] coordinates {};

\nextgroupplot[
  xticklabels={Mean Radius Error},
  ymin=0.00441, ymax=0.04646,
  ytick={0.01,0.02,0.03,0.04},
  yticklabel style={/pgf/number format/fixed,
                    /pgf/number format/precision=3,
                    font=\tiny, xshift=0.4ex}
]
\addplot+[boxplot prepared={
  median=0.021543, lower quartile=0.018517, upper quartile=0.024741,
  lower whisker=0.006317, upper whisker=0.044548
}] coordinates {};

\end{groupplot}
\end{tikzpicture}%
}
\vspace{-25pt}

\caption{Boxplots of Absolute Mean Errors Across all Spherical Primitives.}
\label{fig:boxplots-final}
\end{figure}

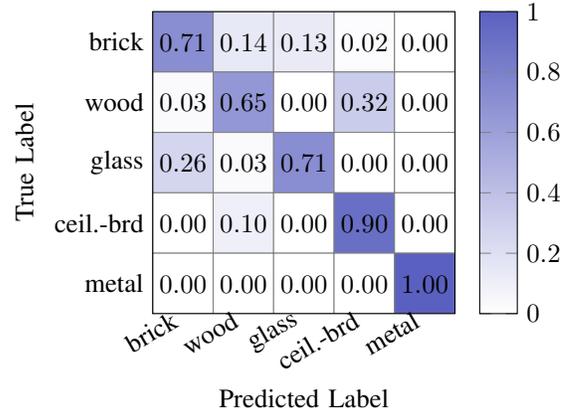
\begin{figure}[t!]
    \centering

    \begin{tikzpicture}
    \begin{axis}[
            width=5.6cm, height=5.6cm,
            colormap={bluewhite}{color=(white) rgb255=(90,96,191)},
            xlabel=Predicted Label,
            xlabel style={yshift=1pt},
            ylabel=True Label,
            ylabel style={yshift=1pt},
            xticklabels={brick, wood, glass, ceil.-brd, metal}, 
            xtick={0,...,4}, 
            xtick style={draw=none},
            yticklabels={brick, wood, glass, ceil.-brd, metal}, 
            ytick={0,...,4}, 
            ytick style={draw=none},
            enlargelimits=false,
            colorbar,
            xticklabel style={
              rotate=30, anchor=east
            },
            nodes near coords={\pgfmathprintnumber[fixed,precision=2,fixed zerofill]\pgfplotspointmeta},
            nodes near coords style={
                yshift=-7pt
            },
        ]
        \addplot[
            matrix plot,
            mesh/cols=5, 
            point meta=explicit,draw=gray
        ] table [meta=C] {
            x y C
            0 0 0.71
            1 0 0.14
            2 0 0.13
            3 0 0.02
            4 0 0.00
            
            0 1 0.03
            1 1 0.65
            2 1 0.00
            3 1 0.32
            4 1 0.00
            
            0 2 0.26
            1 2 0.03
            2 2 0.71
            3 2 0.00
            4 2 0.00
    
            0 3 0.00
            1 3 0.10
            2 3 0.00
            3 3 0.90
            4 3 0.00
    
            0 4 0.00
            1 4 0.00
            2 4 0.00
            3 4 0.00
            4 4 1.00
            
        }; 
    \end{axis}
    \end{tikzpicture}

    \vspace{-11pt}
    
    \caption{Confusion matrix for the material classification task.}
    \label{fig:confusion_matrix}
\end{figure}

\begin{figure*}[ht!]
    \centering

    \begin{minipage}[t]{0.49\linewidth}
        \centering
        \includegraphics[width=\linewidth]{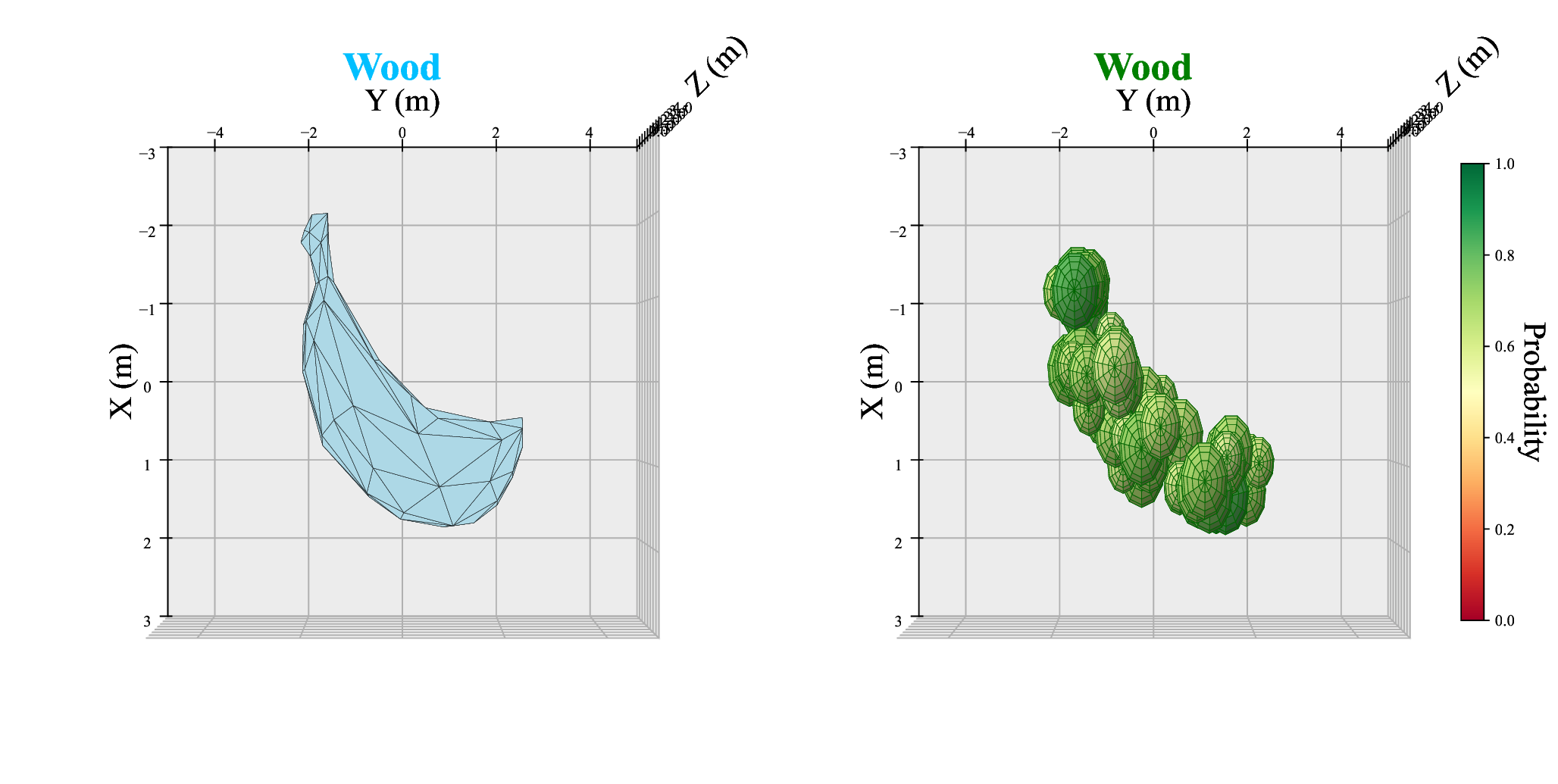}
    \end{minipage}\hfill
    \begin{minipage}[t]{0.49\linewidth}
        \centering
        \includegraphics[width=\linewidth]{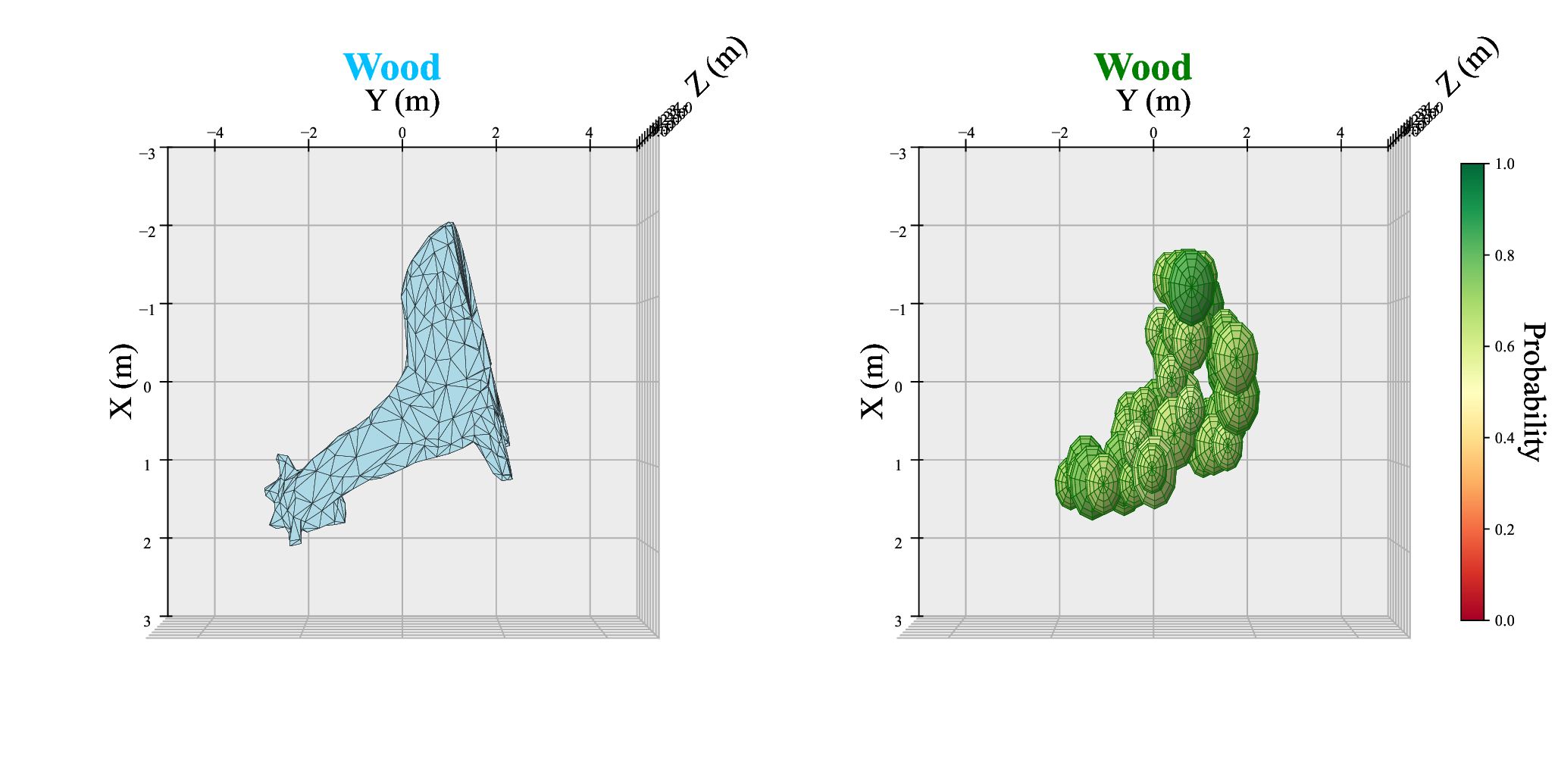}
    \end{minipage}

    \vspace{-25pt}

    \begin{minipage}[t]{0.49\linewidth}
        \centering
        \includegraphics[width=\linewidth]{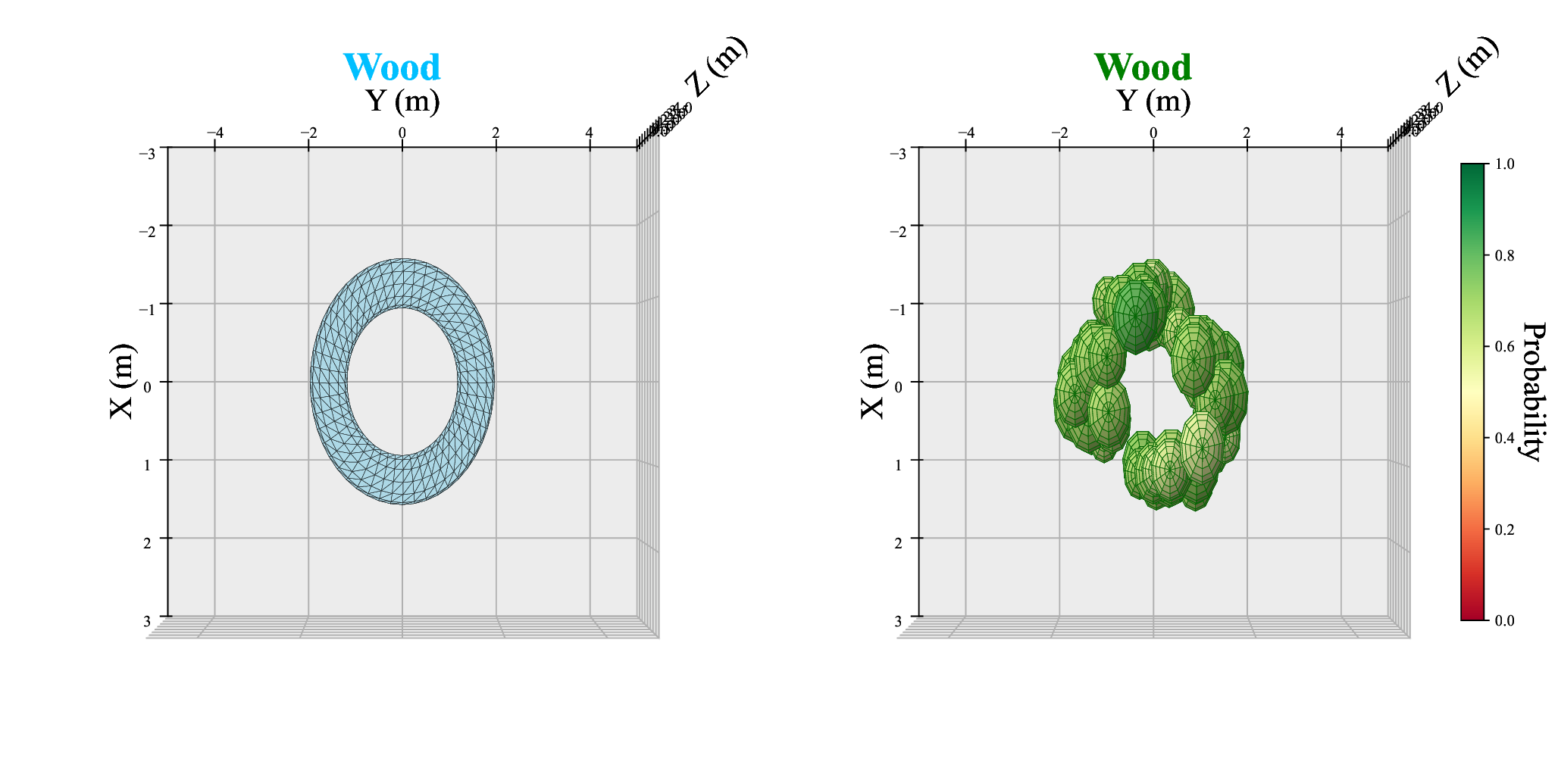}
    \end{minipage}\hfill
    \begin{minipage}[t]{0.49\linewidth}
        \centering
        \includegraphics[width=\linewidth]{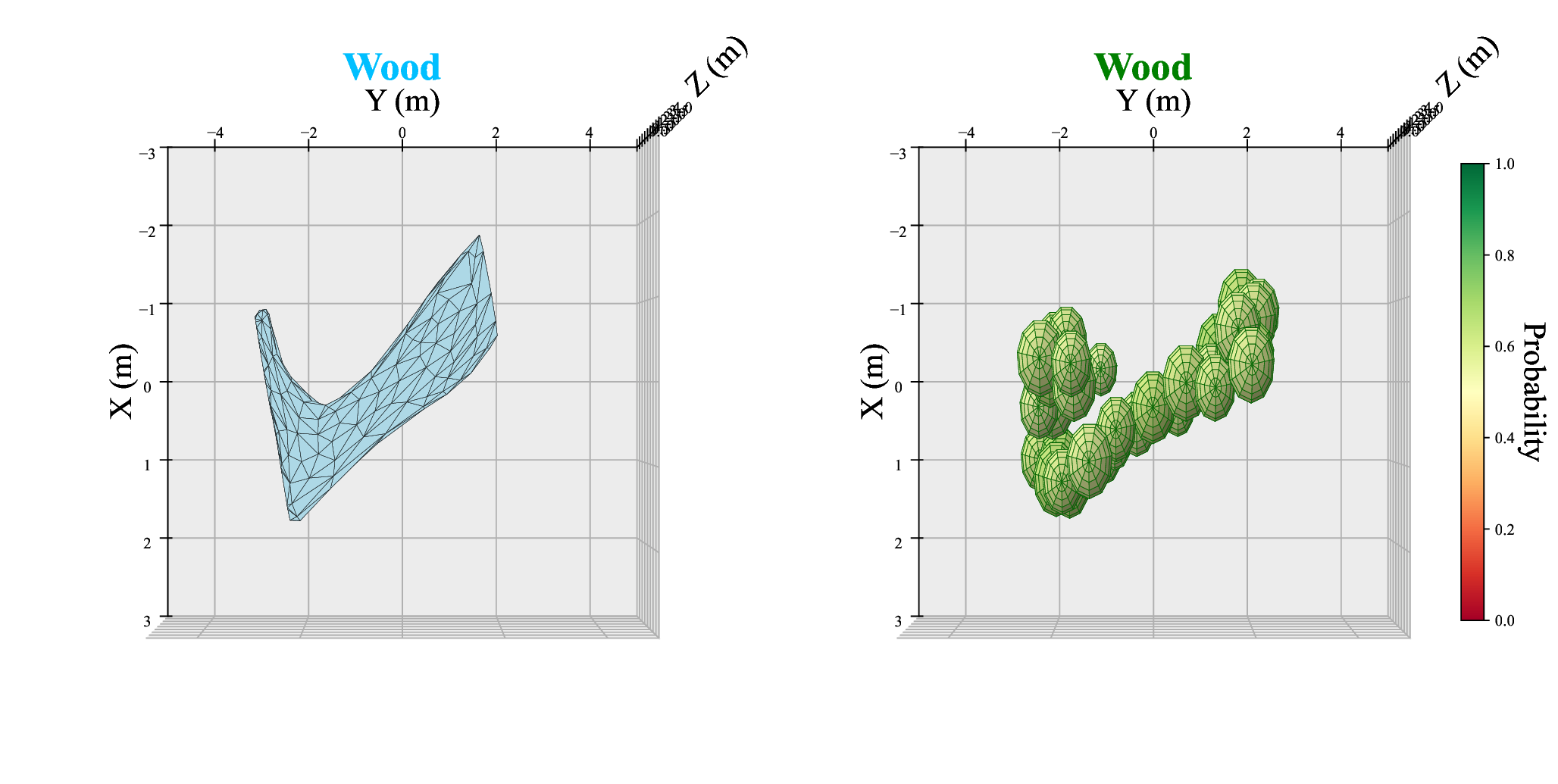}
    \end{minipage}

    \vspace{-25pt}
    
    \caption{Qualitative results across $4$ arbitrary object geometries.}
    \label{fig:res_2x2}
\end{figure*}

Fig.~\ref{fig:boxplots-final} illustrates the distribution of absolute mean errors across all spherical primitives during the testing phase of the DETR-based reconstruction model. The aggregated mean errors over the $12$ spheres are $e_x = 17.48\,\mathrm{cm}$, $e_y = 40.61\,\mathrm{cm}$, $e_z = 3.04\,\mathrm{cm}$, and $e_r = 2.17\,\mathrm{cm}$. Errors remain sub-meter, indicating robust generalization to unseen PWE realizations. Apparent absolute error differences reflect unequal parameter ranges, seen in Table \ref{tab:sim_params}. After normalizing by each span, mean relative errors converge. This suggests that diffuse PWE illumination encodes localization with nearly uniform fidelity across geometric dimensions.



Fig.~\ref{fig:confusion_matrix} illustrates the confusion matrix summarizing the classification performance of the proposed model with respect to spherical primitives material composition. The overall accuracy reaches \(79.35\%\) across all material classes. As can be seen, materials with distinct EM signatures, such as metal, exhibit near-perfect recall due to their high conductivity. Ceiling board follows with a recall of \(90\%\), while glass and brick achieve \(71\%\), and wood remains the most challenging material with \(65\%\). The dominant sources of confusion occur between glass and brick, as well as between the low-permittivity pair of wood and ceiling board. These results are consistent with the EM properties listed in Table~\ref{tab:material_properties}, where the corresponding materials exhibit similar permittivity values that lead to comparable scattering patterns. Nevertheless, further enhancement is anticipated through PWE programming strategies that incorporate polarization-aware control, since polarization responses are inherently more sensitive to material differences.

Finally, Fig.~\ref{fig:res_2x2} presents representative reconstruction results for $4$ arbitrary objects—banana, giraffe, torus, and leg geometries—obtained during the evaluation phase of the proposed framework. Each object is decomposed into an assembly of spherical primitives derived from the DETR predictions using $N=64$ queries and a confidence threshold of $\tau=52\%$. In each subfigure, the left-hand side shows the original object mesh, while the right-hand side depicts its spherical approximation. The reconstructed shapes effectively capture the overall geometry of the targets, confirming the model’s capability to recover complex structures from purely RF-based observations. The banana and leg objects display strong geometric consistency, with the predicted spheres closely following the contour and curvature of the meshes. For the torus, the spherical-primitive approximation captures the overall form but with reduced fidelity, whereas in the giraffe case the body and neck are reconstructed while the head is not. Overall, these results demonstrate that the learned model yields a  physically consistent volumetric representation of arbitrary shapes, aligning with the principles of 3DGS in the RF domain and validating the feasibility of object-level reconstruction within PWEs.

\section{Conclusion} \label{S: Conclusion}
In this work, a programmable RF-sensing framework for three-dimensional object reconstruction was presented, introducing a 3D Gaussian Splatting adaptation to the RF domain. The proposed system combines programmable wireless environment control with a detection transformer architecture to approximate arbitrary objects through material-aware spherical primitives. The obtained results verified effective geometric reconstruction with sub-meter accuracy and a material classification performance of $79.35\%$, confirming the feasibility of data-driven, physically interpretable RF-based volumetric modeling within PWEs.

\section*{Acknowledgments}
This work has been funded by the European Union’s Horizon 2020 research and innovation programs under grant agreement No 101139194 6G Trans-Continental Edge Learning and No 101120779 CYBERSECDOME.

\bibliographystyle{IEEEtran}
\bibliography{bibliography}

\end{document}